%% file: main.tex
\documentclass[10pt,journal,compsoc]{IEEEtran}
\IEEEoverridecommandlockouts
\usepackage{cite}
\usepackage{amsmath,amssymb,amsfonts}
\usepackage{algorithmic}
\usepackage{graphicx}
\usepackage{textcomp}
\usepackage{xcolor}
\usepackage{comment}
\usepackage{subcaption}
\usepackage{listings}
\usepackage{parcolumns}
\usepackage{siunitx}
\usepackage{pgfplots}
\usepackage{nth}

\pgfplotsset{width=8cm,compat=1.9}
\def\BibTeX{{\rm B\kern-.05em{\sc i\kern-.025em b}\kern-.08em
    T\kern-.1667em\lower.7ex\hbox{E}\kern-.125emX}}
\begin{document}

\title{VPIC 2.0: Next Generation \\ Particle-in-Cell Simulations}

\author{Robert Bird,
        Nigel Tan, %Equal contribution to manuscript
        Scott V.~Luedtke, 
        Stephen Lien Harrell,
       Michela Taufer,
       Brian Albright% <-this % stops a space
\IEEEcompsocitemizethanks{
\IEEEcompsocthanksitem Nigel Tan and Michela Taufer are with the University of Tennessee at Knoxville, Knoxville, TN, USA.\hfil\break
E-mail: ntan1@vols.utk.edu, mtaufer@utk.edu.
\IEEEcompsocthanksitem Robert Bird, Scott V.~Luedtke, and Brian Albright are with Los Alamos National Laboratory,Los Alamos, NM, USA.\hfil\break
E-mail: bird@lanl.gov, 	sluedtke@lanl.gov, 	balbright@lanl.gov.
\IEEEcompsocthanksitem Stephen Lien Harrell is with the Texas Advanced Computing Center at the University of Texas at Austin, Austin, TX, USA.\hfil\break
E-mail: sharrell@tacc.utexas.edu.
}% <-this % stops an unwanted space
}

\maketitle

\begin{abstract}

VPIC is a general purpose Particle-in-Cell simulation code for modeling plasma phenomena such as magnetic reconnection, fusion, solar weather, and laser-plasma interaction in three dimensions using large numbers of particles. VPIC's capacity in both fidelity and scale makes it particularly well-suited for plasma research on pre-exascale and exascale platforms. In this paper we demonstrate the unique challenges involved in preparing the VPIC code for operation at exascale, outlining important optimizations to make VPIC efficient on accelerators. Specifically, we show the work undertaken in adapting VPIC to exploit the portability-enabling framework Kokkos and highlight the enhancements to VPIC's modeling capabilities to achieve performance at exascale. We assess the achieved performance-portability trade-off through a suite of studies on nine different varieties of modern pre-exascale hardware. Our performance-portability study includes weak-scaling runs on three of the top ten TOP500 supercomputers, as well as a comparison of low-level system performance of hardware from four different vendors.

\end{abstract}

\markboth{February 17, 2021}{}%

\begin{IEEEkeywords}
Simulation, Portability, Plasma Physics, Particle-in-Cell
\end{IEEEkeywords}

\input{introduction}

\input{background}

\input{vpic}
\input{portability}
\input{performance}

\input{usecases}

\input{related_work}

\input{conclusion}

 \section{Acknowledgments}

 Work performed under the auspices of the U.S. DOE by Triad National
 Security, LLC, and Los Alamos National Laboratory(LANL). This work was supported by the LANL ASC and Experimental Sciences programs. Approved for public release: LA-UR-21-21453. The UTK authors acknowledge the support of LANL under contract \#578735 and IBM through a Shared University Research Award.

 The authors thank John Cazes and Tommy Minyard at the Texas Advanced Computing Center at the University of Texas at Austin,  Preston Smith and Xiao Zhu from the Research Computing department at Purdue University, the Innovative Computing Laboratory at UTK, and Max Katz and the entire team at the NVIDIA Corporation for allocating time for and assisting with the experiments in this paper. Authors Robert Bird and Nigel Tan have made equal contributions to the paper.

 Additionally we thank the Stony Brook Research Computing and Cyberinfrastructure, and the Institute for Advanced Computational Science at Stony Brook University for access to the innovative high-performance Ookami computing system, which was made possible by a \$5M National Science Foundation grant (\#1927880).

\bibliographystyle{IEEEtran}
\bibliography{bibliography}

\section*{Author Biographies}
  
\vskip -2\baselineskip plus -1fil %RFB: This is me committing crimes against Latex, sorry!

\begin{IEEEbiography}
[{\includegraphics[width=1in,height=1.25in,clip,keepaspectratio]{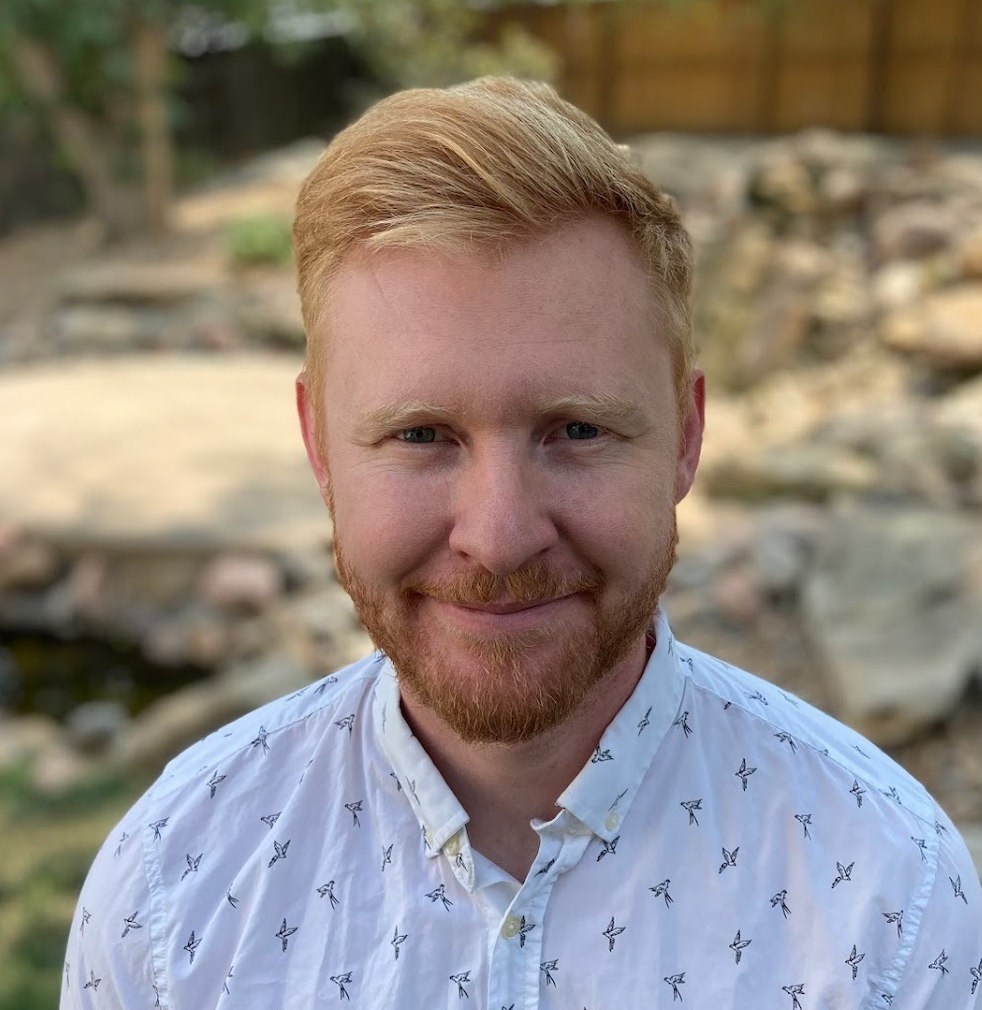}}]
{Robert Bird}
is a Computational Scientist at Los Alamos National Laboratory, in the Applied Computer Science group. His research interests are performance-portability and low-level code optimization. He is the computer science lead for the VPIC project. Dr. Bird received his Ph.D in Computer Science from the University of Warwick, England. 
\end{IEEEbiography}

\vskip -2\baselineskip plus -1fil

\begin{IEEEbiography}
[{\includegraphics[width=1in,height=1.25in,clip,keepaspectratio]{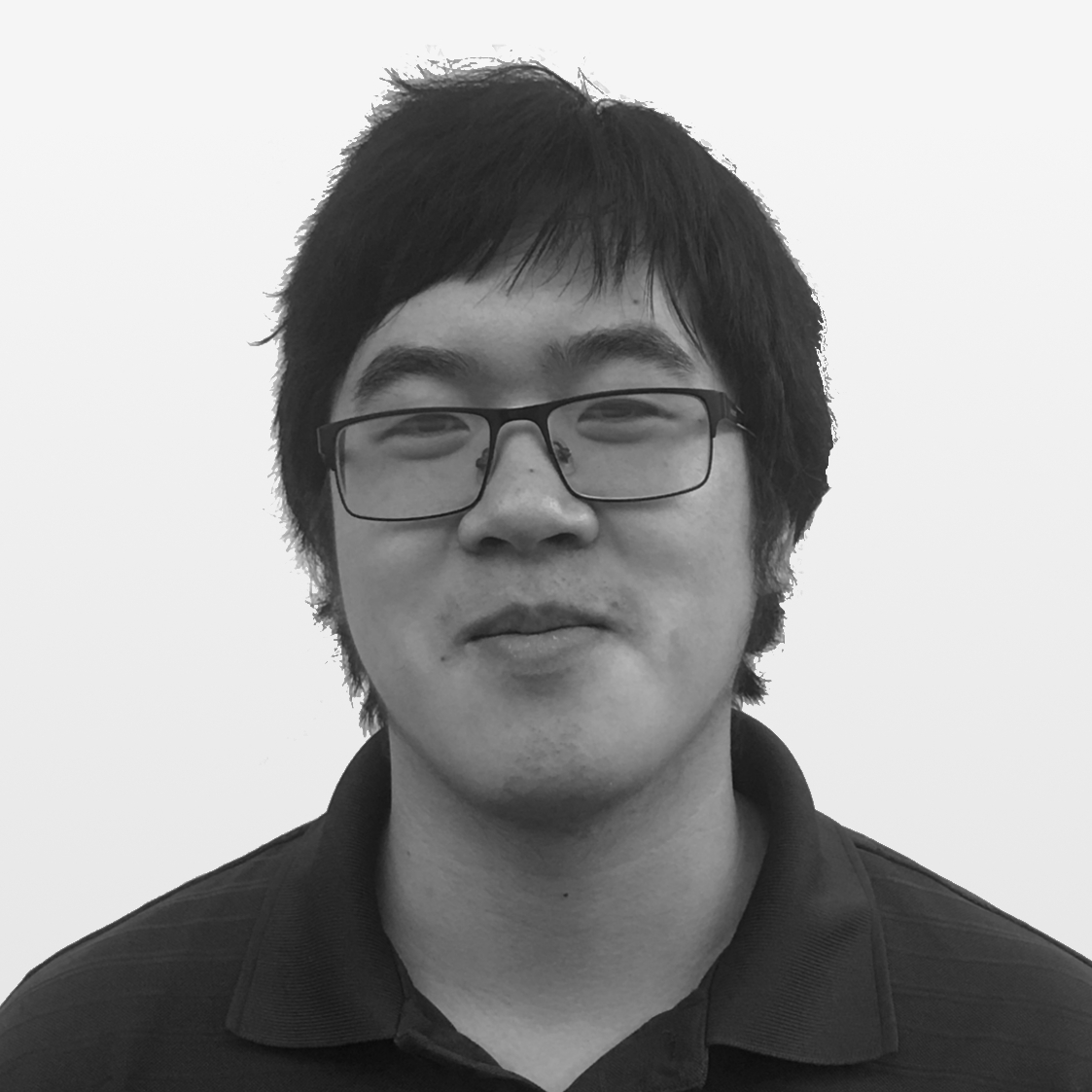}}]
{Nigel Tan}
is a Ph.D. student in Computer Science under Dr. Michela Taufer at the University of Tennessee, Knoxville. He earned his B.S. in both Computer Science and Applied Math at the University of California Merced before earning an M.S. in Computational and Applied Math at Rice University. 
Nigel's research interests lie in high performance computing with an emphasis on performance portability and optimization across multiple architectures.
\end{IEEEbiography}

\vskip -2\baselineskip plus -1fil

\begin{IEEEbiography}
[{\includegraphics[width=1in,height=1.25in,clip,keepaspectratio]{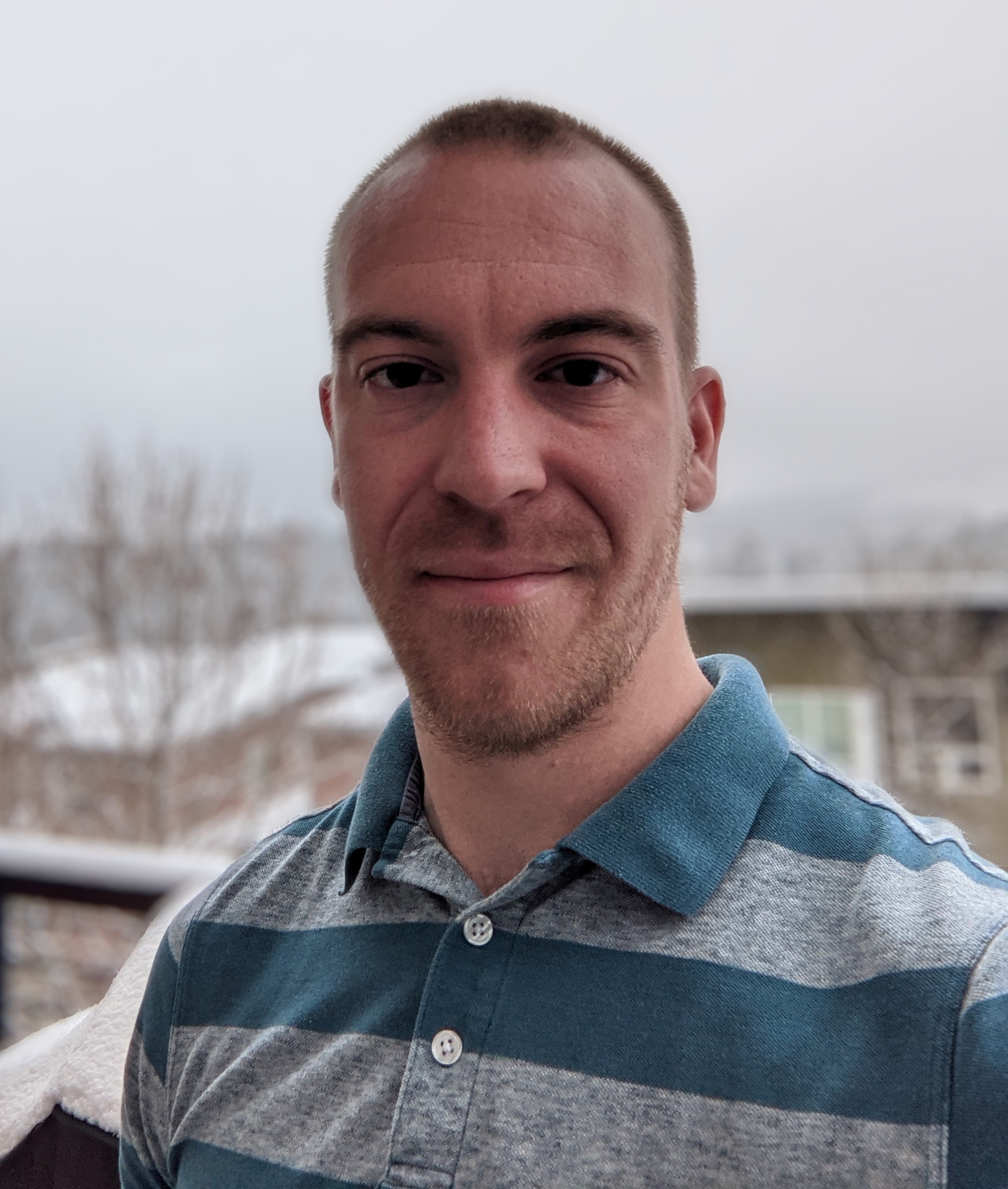}}]
{Scott V.~Luedtke} 
is a Postdoc Research Associate at Los Alamos National Laboratory in the Plasma Theory and Applications group.  His research interests include high energy density physics, high intensity short-pulse laser-plasma interactions, and large-scale physical simulation.  Dr.~Luedtke received his Ph.D.\ in physics from the University of Texas at Austin in 2020.
\end{IEEEbiography}

\vskip -2\baselineskip plus -1fil

\begin{IEEEbiography}
[{\includegraphics[width=1in,height=1.25in,clip,keepaspectratio]{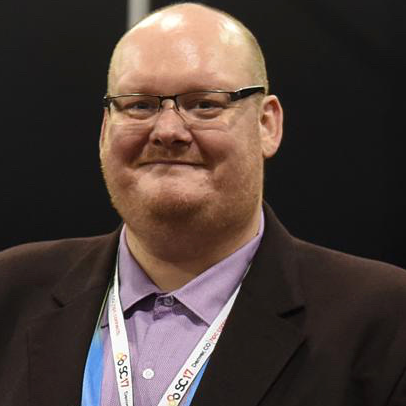}}]
{Stephen Lien Harrell}
is an Engineering Scientist at the Texas Advanced Computing Center in the HPC Performance and Architectures group. His research interests include performance portability, performance modelling, benchmarking and HPC metric capture. Before his current appointment Stephen worked as an HPC System Administrator and HPC Support Staff for twelve years and received his bachelors degree in Computer Science at Purdue University. 
\end{IEEEbiography}

\vskip -2\baselineskip plus -1fil

\begin{IEEEbiography}
[{\includegraphics[width=1in,height=1.25in,clip,keepaspectratio]{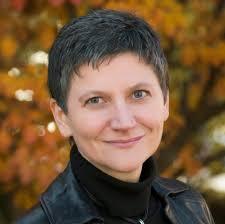}}]
{Michela Taufer}
(Senior Member, IEEE) is an ACM Distinguished Scientist and holds the Jack Dongarra professorship in high performance computing with the Department of Electrical Engineering and Computer Science, University of Tennessee Knoxville. Her research interests include high-performance computing, volunteer computing, scientific applications, scheduling and reproducibility challenges, and in situ data analytics. Dr. Taufer received her Ph.D. in Computer Science from the Swiss Federal Institute of Technology (ETH) in 2002.
\end{IEEEbiography}

\vskip -2\baselineskip plus -1fil

\begin{IEEEbiography}
[{\includegraphics[width=1in,height=1.25in,clip,keepaspectratio]{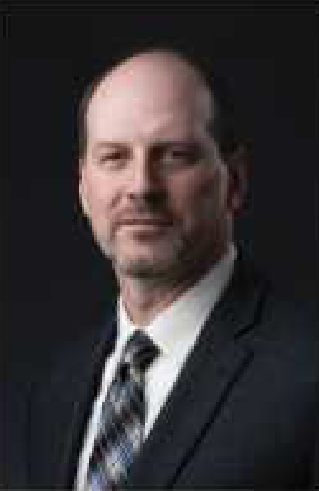}}]
{Brian Albright}
(Senior Member, IEEE) is a Fellow of the American Physical Society and a Laboratory Fellow in the X-Theoretical Design Division of Los Alamos National Laboratory. His research interests include plasma and high energy density physics, high performance computing, numerical methods, and defense applications. Dr. Albright received his Ph.D. in physics from UCLA in 1998. 
\end{IEEEbiography}

\end{document}

%% file: introduction.tex
\section{Introduction}

Many plasma physics phenomena can only be understood through the use of kinetic Particle-in-Cell (PIC) simulations. 
VPIC~\cite{bowers20080,bowers2008ultrahigh,bowers2009advances} has been at the forefront of this research to study plasma physics phenomena including magnetic reconnection, fusion, solar weather, and laser-plasma interaction.
VPIC was deployed to perform high performance multi-trillion particle simulations on over 2 million MPI processes on the Trinity supercomputer and other computing clusters at Los Alamos National Laboratory~\cite{bowers2009advances, stark2018detailed}.

The original VPIC code (referred to throughout as VPIC 1.0) was built to minimize data movement and maximize performance; given a new HPC platform, the ad-hoc re-engineering of the code had guaranteed the continuation of performance. 
However, the increasing heterogeneity of emerging exascale platforms makes it no longer practical to optimize the VPIC code for every available compute platform.  A new multi-architectural VPIC code is needed to ensure portability and performance across exascale platforms, but there are three challenges to building this code. First, the build system complexity grows significantly with the number of architectures supported. Second, the code must allow usage of native hardware focused APIs and libraries. 
Hardware vendors and library developers put significant effort into optimizing common operations and function calls.
Not allowing access to existing optimized libraries creates more work for the developers and reduces performance.
The third, and probably the most tangible and difficult, challenge is loss in performance. Portable code comes at the cost of performance due to differences in parallelism approaches among platforms. Writing portable code requires forcing a single parallelism model onto many platforms. The mismatch between model and platform leads to a loss in performance.

There are multiple methods to overcome these challenges and to create a multi-architectural VPIC code while reducing long-term maintenance. These include pragma-based directives such as OpenMP  and template-based library approaches such as  Kokkos~\cite{CarterEdwards20143202} and Raja~\cite{hornung2014raja}. Template-based library methods are particularly powerful because the can reduce the amount of code needed to support multi-architectural codes. Furthermore, frameworks like Kokkos and Raja can make architectural code-branching decisions at compile time while minimizing the architecture-dependent code required by the application developers. In this paper, we address the three challenges listed above to transformed the original VPIC code (VPIC 1.0), a CPU-only, platform-specific optimized code,  into VPIC 2.0; a multi-architectural, portable code, with the support of the Kokkos abstraction layer. The single VPIC source code runs across a diverse range of modern hardware.

Transforming VPIC into a high performance portable code capable of running on emerging heterogeneous platforms requires several changes. Using Kokkos effectively involves changes to both VPIC's data structures and user interfaces. VPIC 2.0 alters data storage to leverage the benefits of Kokkos Views (advanced multidimensional arrays). Using Views for storage allows us to automatically adjust the memory layout to best match the target hardware. Extending the user interface is required to maintain backwards compatibility and improve the performance of simulation diagnostics. Other transformations  that we introduce focus on further improving the portability and performance of VPIC 2.0 through code restructuring and optimizations. 

VPIC 2.0 scalability and portability is tested across nine platforms including the Sierra, Frontera, and Selene supercomputers. We demonstrate excellent weak scaling, especially the GPU-based platforms where we see less than 10\% slowdown between 64 and up to 7,200 GPUs. The rise in GPU cache size opens up intriguing possibilities for super-linear speedup at tangible scientific scales, as shown in our strong scaling results. Our portability study demonstrates the effectiveness of VPIC 2.0 across all nine platforms and highlights both the challenges and strengths of portable code.
VPIC 2.0 is easy to adapt and run on emerging hardware platforms with limited manual work. The portability of VPIC 2.0 enables the use of cutting edge accelerators for greater performance and facilitates longer simulations, more advance diagnostics, and faster turn around time.
The contributions of this paper are as follows:
\begin{itemize}
    \item An effective approach for performance portability
    \item The performance and portability study on nine platforms
    \item The study of weak and strong scaling across CPU and GPU platforms (including 3 top 10 machines)
    \item The analysis of the costs and benefits to portable code
    \item Insights on the impacts of VPIC 2.0 on future scientific discovery
\end{itemize}

The rest of this paper is structure as follows: Sec~\ref{sec2} presents key changes in modern computing and provides a overview of Kokkos; Sec~\ref{sec3} introduces the physics behind VPIC; Sec.~\ref{sec4} presents our transformation to the VPIC code needed to achieve portability; Sec~\ref{sec5} shows performance and scalability results; Sec~\ref{sec6} describes the impact of the newly release VPIC on scientific discovery at exascale; Sec~\ref{sec7} cites related work; and Sec~\ref{sec8} summarizes this work.

%% file: background.tex
\section{Background}
\label{sec2}

\subsection{Modern Heterogeneous Computing}

Supercomputing has a rich history of exploring and exploiting different aspects of parallelism. From the first vector supercomputers in the 1970s, to the massive clusters of computers of today, accelerators have been used to offload time-consuming tasks from the primary compute unit(s). This was first seen with specific-purpose co-processors~\cite{gu2008explicit, cher2008cell}, and later evolved to include generic purpose accelerator hardware~\cite{surmin2014particle} and GPUs~\cite{kindratenko2009gpu} as the cornerstone of accelerator based computing. 

Accelerators today provide a highly parallelized compute capacity. We have seen a few major approaches in accelerators in the last 10 years. The first, the GPGPU which uses hundreds of graphics cores that often only compute at single or half precision. The second modern accelerator type contains many general purpose cores which are generally slower than the primary compute unit. Some examples of this type are the Intel Knights Corner~\cite{chrysos2014intel} and Sunway SW26010~\cite{LIN2018128} accelerators. Finally, while a large amount of SIMD capability now sits on the package of most modern CPUs, a third group of accelerators exists that provides additional vector capability, such as the NEC Vector Engine~\cite{komatsu2018performance}. 

These characteristics and the accompanying execution models of modern accelerators create many challenges for maintaining and improving performance, which is the primary motivation for using accelerators in high performance computing. Porting an existing code base to a new accelerator architecture while maintaining performance can have many facets including: novel memory models; explicit concerns around data locality and data movement; unfamiliar execution models; and changes in the scale of parallelism. Any of these issues can be core to maintaining performance through an architecture change. Further compounding this complexity is that once a code base is ported to a new platform and is performant, there may now be a new branch or code variant to maintain.

\subsection{Kokkos Overview}

Kokkos \cite{CarterEdwards20143202} is a suite of libraries and abstractions for developing high performance C++ applications that are portable across many architectures, including all major CPUs and accelerators. Kokkos achieves high performance and portability by mapping its own programming model to various native backend parallel programming languages such as CUDA or HIP. 

There are three major elements of the Kokkos ecosystem:  the Kokkos kernels libraries, the Kokkos tools interface and the Kokkos core programming model. The Kokkos kernels libraries provide common linear algebra (e.g., BLAS, SparseBLAS) and graph algorithms. Kokkos tools are comprised of an interface and a set of utilities that connect to the Kokkos runtime and enable lightweight debugging and profiling. The Kokkos programming model enables developers to control memory layout, placement of data, parallel execution, and execution devices. Kokkos maps specified data characteristics and execution policies to the target hardware backends using a set of rules for automatically determining different hardware specific parameters, removing the user from the mapping task. 

\begin{figure}[!ht]
    \centering
    \includegraphics[width=0.45\textwidth]{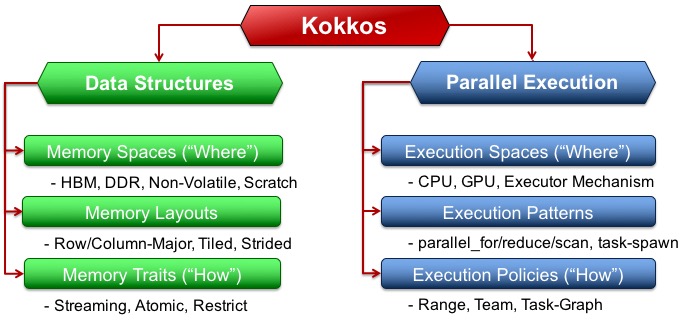}
    \caption{Core Kokkos abstractions for ensuring performance portability~\cite{trott2017solving}. }
    \label{fig:kokkos}
\end{figure}

The Kokkos programming model abstractions manage common data structures and parallel execution as seen in Fig.~\ref{fig:kokkos}.
Kokkos data structures have three key extensions defining the memory space, layout, and traits. The memory space determines where the memory is allocated. 
This abstraction enables the use of different memory technologies (e.g., High Bandwidth Memory or HBM, DDR SDRAM, and Non-Volatile Memory). Memory layout controls the order in which memory is allocated and stored, affecting the memory access pattern. Kokkos includes the most common layouts (e.g., row/column major, strided, and tiled). These layouts are important for maximizing performance and interoperability with various linear algebra libraries. Kokkos also supports various memory traits that affect how memory is accessed (e.g., streaming, atomic, restricted) and offer the users fine grained control to be used to expose additional information or optimization opportunities. 
Kokkos Views are the most common data structure used in VPIC. Views are reference counted, multi-dimensional arrays with optional memory traits to control memory access. These memory traits, combined with hardware settings determined at compile time, enable Kokkos applications to operate on significantly different hardware platforms with a single code path needed to be written by the developer. For example, GPU applications generally require memory to be setup on the host and copied to the device. Kokkos addresses this pattern by having a HostMirror View that acts as the host copy of a device View. Developers can use a \texttt{deep\_copy} operation to move data between Views, including between a host and a device. If the target device is the same as the host, in the case of a CPU only build,  any unnecessary \texttt{deep\_copy} will change into a \texttt{noop}.

Kokkos's parallel executions are defined via execution spaces, patterns, and policies. An execution space defines where an operation is to be run (e.g., CPU or GPU and the execution mechanisms). The pattern of execution (or parallel dispatch) defines the parallel operation to execute (e.g., \texttt{parallel\_for}, reduce, or scan, and \texttt{task\_spawn}). Execution policies define how the operation should be executed (e.g., over a range of indices) with a team of threads, or using a task-graph. Given an application, when it is time to define its parallel execution, Kokkos can either automatically choose hardware specific settings or allow developers to use their own settings for finer performance tuning.

%% file: vpic.tex
\section{Vector Particle-In-Cell (VPIC) Overview} 
\label{sec3}

The Particle-In-Cell method is among the most widely used computational tools in plasma physics. Being fully kinetic, PIC codes are used when fluid approximations to plasma physics fail.
With its first principles approach to the physics and fast explicit time stepping, PIC codes have been used to simulate plasmas ranging from micron- and femtosecond-scale short-pulse laser-plasma interactions~\cite{yin2007monoenergetic,yin2011three} to lightminute- and day-scale supermassive black hole accretion disks~\cite{levinson2018particle,hoshino2015angular}.
Other uses include: laser-plasma instability~\cite{albright2016multi} and cross-beam energy transfer~\cite{yin2019saturation} simulations for Inertial confinement fusion; magnetic flux ropes in the heliosphere~\cite{du2018plasma}; magnetic reconnection in laboratory plasmas~\cite{yamada2014conversion};
and Earth's magnetosphere~\cite{chen2016electron}.

The PIC method solves the Maxwell-Boltzmann system of equations 

\begin{subequations}
    \label{maxwell}
\begin{align}
    {\nabla} \cdot \vec{E} &= \frac{1}{\epsilon_0}\rho, \label{gauss}\\
    {\nabla} \times \vec{E} &= -\frac{\partial\vec{B}}{\partial t}, \label{faraday}\\
    {\nabla} \cdot \vec{B} &= 0, \label{monopole}\\
    {\nabla} \times \vec{B} &= \mu_0 \vec{J} + \mu_0 \epsilon_0 \frac{\partial\vec{E}}{\partial t}, \label{ampere}
\end{align}
\end{subequations}

\begin{align}
    \partial_t f_s + \vec{v} \cdot \nabla_x f_s +  q_s \left[\vec{E} + \vec{v} \times \vec{B} \right] \cdot \nabla_p f_s  = C_s(f),
\label{boltz}
\end{align}
where $\nabla$ is the del operator, $\vec{E}$ is the electric field, $\vec{B}$ is the magnetic or B field (in Tesla), $\rho$ is the charge density, $\vec{J}$ is the current density, $\epsilon_0$ and $\mu_0$ are the vacuum permittivity and permeability, respectively, $C_s(f)$ is the collision operator, and $f_s = f_s(\vec{x}, \vec{v}, t) = \langle \mathcal{F}_s(\vec{x}, \vec{v}, t)\rangle$ is the plasma distribution function, where the brackets denote the ensemble average and $\mathcal{F}_s$ is the microscopic phase-space distribution function, the subscript $s$ denotes the species, $\vec{x}$ is position, $\vec{v}$ is velocity, and $\vec{p}$ is momentum.

Conceptually, the PIC method solves the electromagnetic fields on a grid and moves particles continuously through space. The particles move in response to the fields according to the Lorentz force law, and deposit current on the grid. They do not directly interact with each other, unless collisions are modeled. The fields evolve according to Maxwell's equations, Eqs.~\ref{maxwell}, including the current from the particles.

The plasma distribution function is represented by quasiparticles, also called simulation particles or macroparticles. Quasiparticles are single particles of a particular species in the simulation that represent a number of real particles given by their weight, $w$.
The PIC method thus samples the smooth distribution function $f_s$, which is the ensemble average of the ``ground truth" microscopic $\mathcal F_s$.
Typically, $w \gg 1$.

Furthermore, quasiparticles have a shape (i.e., a distribution of mass charge over some spatial extent).
For a given particle shape, $\phi$, the distribution function is represented in PIC as
\begin{equation}
    f_s(\vec{x}, \vec{v}, t) = \sum_{i=1}^{N_s } w_i \phi(\vec{x} - \vec{x}_i(t)) \,\delta(\vec{v} - \vec{v}_i(t)).
\end{equation}
Typically, the particle shape functions are B-splines:
\begin{equation}
b_0(\xi) = \left\{ \begin{array}{ll}
1 & \textrm{if $|\xi| < 1/2$}\\
0 & \textrm{otherwise}
\end{array} \right . ,
\end{equation}
\begin{equation}
b_{n+1} = \int_{-\infty}^\infty b_0(\xi - \xi')b_n(\xi-\xi')\,d\xi' ,
\end{equation}
where $\xi$ is the distance to the quasiparticle center normalized to the grid spacing.
Zeroth order is called top-hat shape, and first is triangle.
The 3D shape function is typically the tensor product of the 1D shape functions in each dimension.
The field experienced by a quasiparticle with a finite spatial extent is
\begin{equation}
    \vec{E}_i = \int \vec{E}\, \phi(\vec{x}-\vec{x}_i) d\vec{x},
\end{equation}
\begin{equation}
    \vec{B}_i = \int \vec{B}\, \phi(\vec{x}-\vec{x}_i) d\vec{x}.
\end{equation}
Since the fields are defined on the grid, evaluating these integrals requires an interpolation scheme \cite{villasenor1992rigorous,esirkepov2001exact}, the details of which can be messy.

All PIC codes at their core consist of two  coupled solvers: the particle pusher and the field solver.
The particle pusher moves plasma particles freely through space based on surrounding EM fields and calculates currents arising from this motion. The field solver solves Maxwell's equations on a fixed grid, accounting for the currents from the particles. These core algorithms fully reproduce the behavior of collisionless, relativistic, kinetic plasmas. Additional physics, such as collisions and special boundary conditions, can be added with additional physics modules.

Maxwell's equations are solved using the widely used finite-difference time-domain (FDTD) method \cite{fdtd} on the Yee staggered grid \cite{yee}.
The staggered Yee grid allows for easily implemented centered, second order accurate derivatives.
The particle pusher solves the relativistic equations of motion
\begin{equation}
\frac{d\vec{x}}{dt} = \vec{v},
\end{equation}
and
\begin{equation}
\frac{d(\gamma\vec{v})}{dt} = \frac{q}{m}(\vec{E + v \times B}).
\end{equation}

VPIC uses the Boris pusher \cite{birdsall2004}, a second-order accurate pusher that employs the leapfrog method, slightly modified to avoid cyclotron motion aliasing, similar to the method used in \cite{blahovec20003}.
The Boris pusher splits the equations of motion into an acceleration from the $\vec{E}$ field and a rotation about the $\vec{B}$ field.
A particle is first pushed a half-step, based on its previous momentum.
The fields are interpolated to this half-step position, the momentum is updated with these fields, and the particle is moved the full step.
Finally, the particle is moved to the three-halves position to calculate the current used by the field solver, and the three-halves position is discarded.

More formally, the difference equations are
\begin{equation}
    \frac{\vec{x}^{n+\frac{1}{2}} - \vec{x}^{n-\frac{1}{2}}}{\Delta t} = \vec{v}^n,
\end{equation}
and
\begin{equation}
    \frac{\vec{p}^{n+1} - \vec{p}^{n}}{\Delta t} = q (\vec{E}^{n+\frac{1}{2}} + \vec{v}^{n+\frac{1}{2}} \times \vec{B}^{n+\frac{1}{2}}) ,
\end{equation}
where $\vec{v} = \vec{p}/(m\gamma)$, and
\begin{equation}
    \vec{v}^{n+\frac{1}{2}} = \frac{\vec{p}^{n} + \vec{p}^{n+1}}{2m\gamma^{n+\frac{1}{2}}}.
\end{equation}

%% file: portability.tex
\section{VPIC for Heterogeneous Platforms} 
\label{sec4}

\subsection{Data Structure and Interface Transformations}

Two components of the VPIC code need to be redesigned in order to deploy the Kokkos abstraction layer effectively: (1) the VPIC data structure elements and (2) the user interface.

\subsubsection{VPIC Data Structure Elements}

VPIC operates by defining a simulation space divided into a grid of cells, and modeling particle movement across these cells. In other words, particles are distributed across an n-dimensional (n-D) space that is decomposed into a n-D grid. Each simulated particle is a macroparticle with a defined weight (i.e., the number of real particles modeled by each macroparticle). There are four key data elements in VPIC that must be moved to Kokkos Views to ensure portability. They are the macroparticles, electromagnetic (EM) fields, interpolators, and current accumulators. Fig.~\ref{fig:general_macroparticle} describes the original C++ definition of macroparticle storage in VPIC 1.0 before the Kokkos port. Each macroparticle is a 32-byte structure (i.e., 3 floats for voxel position $dx$, $dy$, and $dz$, 3 floats for normalized momentum, $ux$, $uy$, $uz$, 1 float for weight, and 1 integer for cell index). 
\begin{figure}[!ht]
    \centering
    \begin{subfigure}[t]{0.45\textwidth}
    \centering
\begin{lstlisting}[frame=tlrb,language=C++,basicstyle=\scriptsize]{Name}
typedef struct Particle {
    float dx; // Position in (x,y,z) dimensions
    float dy; 
    float dz;
    float ux; // Momentum in (x,y,z) dimensions
    float uy; 
    float uz;
    float w;  // Weight of simulated particle
    int i;    // Index of cell where particle resides
};
Particle particles[N]; // Particle storage
particles[0].dx; // Access dx from particle 0

\end{lstlisting}
    \caption{Original VPIC macroparticle.}
    \label{fig:general_macroparticle}
\end{subfigure}\hfill
\\
\begin{subfigure}[t]{0.45\textwidth}
    \centering
\begin{lstlisting}[frame=tlrb,language=C++,basicstyle=\scriptsize]{Name}
View<float*[7]> particles(N); // Particle data
namespace particle_var { 
    enum p_v { // Particle member enum for clean access
        dx=0,  // Position in (x,y,z) dimensions
        dy,
        dz,
        ux,    // Momentum in (x,y,z) dimensions
        uy,
        uz,
        w,     // Weight of simulated particle
    }; };
View<int*> particle_indices(N); // Particle indices
// Access dx from particle 0
particles(0, particle_var::dx);
\end{lstlisting}
    \caption{VPIC macroparticle using a Kokkos View.}
    \label{fig:kokkos_macroparticle}
\end{subfigure}
    \caption{Code changes for storing and accessing particle data using Kokkos.  \label{fig:kokkos_storage}
    }
\end{figure}

The EM fields are represented in the original C++ VPIC 1.0 code as a 80-byte structure (i.e., 3 floats for the $\vec{E}$ field, 1 float for the divergence of $\vec{E}$ error, 3 floats for the $\vec{B}$ field, 1 float for the divergence of $\vec{B}$ error, 3 floats for free current, 1 float for charge density, 3 floats for the transverse current adjustment field, 1 float for the bound charge density, and 8 short integers for identifying the various materials at different places in the grid); interpolators are represented as  72-byte structure (i.e., 18 floats for the various interpolator values); and current accumulators are represented as 48-byte structure (i.e., 4 floats for the current in the $x$ direction, 4 floats for the $y$ direction, and 4 floats for the $z$ direction).  

\begin{figure}[!ht]
    \centering
    \includegraphics[width=0.45\textwidth]{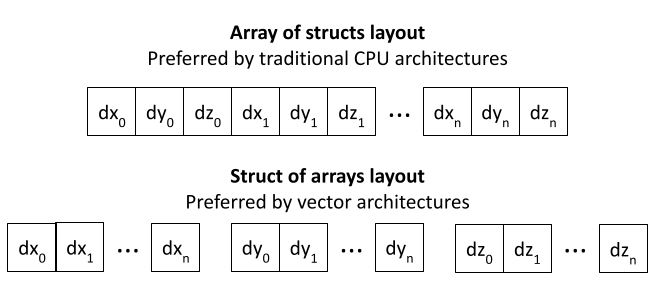}
    \caption{Comparison of Struct of Arrays (SoA) and Array of Structs (AoS) memory layouts for storage position in 3D cartesian coordinates. 
    }
    \label{fig:SoAvAoS}
\end{figure}

Porting VPIC to use Kokkos requires that we store the data structures in Kokkos Views. Care must be taken during the conversion to Kokkos Views because the conversion approach directly affects the memory layout. Memory layout is important for performance, with different hardware performing better with different layouts and parallelism techniques (i.e., thread parallelism and vector parallelism)~\cite{strzodka2012abstraction}. In general there are two major memory layouts, array of structs (AoS) and struct of arrays (SoA). The AoS stores multiple instances of a \texttt{struct} in an array such that they are contiguous in memory. The SoA layout, on the other hand, stores multiple instances of a \texttt{struct} in a single \texttt{struct} with arrays for each structure member. Each member stores values in a contiguous array. Fig.~\ref{fig:SoAvAoS} demonstrates the differences between the layouts using particle position coordinates as an example. The AoS layout has each coordinate \texttt{(dx,dy,dz)} in memory one after the other. The SoA layout has separate arrays for each dimension. The AoS layout is favored by traditional CPU hardware. The array is easily broken into chunks that each thread can operate on independently. Cache efficiency on CPUs also improves with the AoS layout due to the higher likelihood of fitting an entire \texttt{struct} in a single cache line. On the other hand, the SoA layout is preferred for hardware with small caches such as GPUs. On GPUs, many threads execute the same instruction. The small cache size and high memory access latency makes each memory operation very costly. The SoA layout ensures that memory is accessed in a contiguous manner. Accessing contiguous blocks of memory helps the GPU use its full memory bandwidth.

Simply replacing arrays with Views may be inefficient as this decision limits memory layout and decreases performance when, for example, running VPIC on GPUs. Swapping arrays for Views restricts the data to a CPU favoring AoS layout. 
Instead of swapping, we use a 2D View to store data such that the layout is automatically chosen based on the target hardware (i.e., CPU or GPU). Fig.~\ref{fig:kokkos_storage} shows the application of this principle when moving VPICs particle data to a Kokkos View. The original VPIC macroparticle code (VPIC 1.0) in Fig.~\ref{fig:general_macroparticle} stores particles in an array of particles. Fig.~\ref{fig:kokkos_macroparticle} shows how the data is stored and accessed in a 2D Kokkos View in VPIC 2.0. The first and second index determine which particle and member variable to access respectively.

While converting structures to 2D Views, we consider two types of structures: structures with all members having the same member type and structures with multiple member types. Data with structures containing a single member type (e.g., only float) are simply moved to a 2D Kokkos View where the first dimension determines the View size while the second dimension corresponds to the different \texttt{struct} members. For data with structures containing multiple member types, such as the the macroparticle in Fig.~\ref{fig:general_macroparticle}, members of a given type are gathered in their own View and multiple Views are deployed. Fig.~\ref{fig:kokkos_macroparticle} shows how the particle cell index is separated into its own View due to having a different type than the other macroparticle items. Table~\ref{tab:data_structure_transformations} summarizes the types of members and Views for the four key VPIC data elements.
\begin{table}
    \centering
    \begin{tabular}{|l|c|l|}
        \hline
        Data Structure & Member Types & Transformation \\
        \hline
        Particles & \texttt{float} and \texttt{int} & Multiple Views\\
        EM Fields & \texttt{float} and \texttt{short int} & Multiple Views\\
        Interpolators & \texttt{float} & Single View\\
        Accumulators & \texttt{float} & Single View\\
        \hline
    \end{tabular}
    \caption{Member types and necessary code transformations for the four key data structures. Heterogenous data requires separate Kokkos Views for each member type. Homogeneous structures only need a single Kokkos View.}
    \label{tab:data_structure_transformations}
\end{table}

For convenience, we define an \texttt{enum} with enumerals matching the original \texttt{struct} fields. An \texttt{enum} provides a clear definition for accessing \texttt{struct} fields stored in Views. Fig.~\ref{fig:kokkos_storage} shows the difference in code for the particles. In Fig.~\ref{fig:general_macroparticle} the particle is defined using plain C++. Transforming the code with Kokkos results in the particle definition in Fig.~\ref{fig:kokkos_macroparticle}. Both the declaration and data access code for particle storage is similar to the original VPIC 1.0 code from the . This conversion approach provides a single interface to managing data across multiple platforms and allows domain experts to focus on scientific discovery.

\subsubsection{User Interface}

VPIC has several functions that allow the user to collect diagnostics and adjust the simulation. Users define their desired adjustments and diagnostics in C++ along with the frequency at which VPIC should call the functions. Porting VPIC to Kokkos requires changes to the user interface for backwards compatibility and performance. 

VPIC data is stored in Kokkos Views with an explicit separation between host and device data. This  separation is a requirement for parallel execution and needs changes to the code to maintain backwards compatibility with existing VPIC decks. The existing interface assumes all operations are done on the old data structures. Thus data must be copied between the old structures and Kokkos Views. The explicit split between host and device data requires that both structures are synchronized before and after any user diagnostics or adjustments. As a result, data must be copied from the device to the host and copied again to the original VPIC data structures. We  add a set of flags for controlling data movement for user defined functions. By default, VPIC automatically handles data movement such that legacy codes run without changes.

Shuffling data among the host, device, and original structures is costly. Users can avoid the performance impact from moving data by writing their user functions with Kokkos. Functions written using the Kokkos API can directly operate on the data regardless of where it resides. Thus, unnecessary data movement is avoided. This feature also enables users to easily paralellize their code and integrate it into VPIC seamlessly. Under these circumstances, users need only to write their functions and set the appropriate flags.

\subsection{Other Code Transformations}

We perform additional code re-engineering to further improve portability and performance. 

\subsubsection{Improving Portability}

There are two changes to the VPIC code that enable broader portability across platforms. The first, we move away from the vectorization formulation heavily deployed in VPIC 1.0. Second, we replace the explicit data duplication used for scatter-reduction patterns with the more portable ScatterView container provided by Kokkos.

The original VPIC 1.0 code was designed to exploit vectorization with explicit SIMD code. While highly performing on CPUs, the formulation does not fit the GPU programming model and adds significant complexity. We re-engineer the code to match the Kokkos model instead; such a model generalizes well across both CPUs and GPUs. As the Kokkos project develops, we will improve the code to better expose vector parallelism.

A ScatterView is a helpful structure for concurrent reductions. Concurrent updates are implemented in two ways depending on the target hardware. On CPUs, each thread duplicates the data and operates locally before collecting updates from all threads and updating the data. This is done to avoid slow atomic operations and improve cache performance. On GPUs, caches are smaller and atomic operations are faster so updates are done with atomics. Electircal current accumulation in VPIC uses arrays of accumulators so that each thread can collect contributions independently before updating the EM fields. A ScatterView accomplishes the same goal, so we replace the accumulator arrays with a ScatterView that updates the current directly. 

\subsubsection{Improving Performance}

For improving performance on GPU systems we integrate four optimizations in VPIC: restructuring code to keep data resident in the GPU; exploiting hierarchical parallelism; changing particle sorting order; and specializing the reduction code.

Data movement in Kokkos is done explicitly. In our first optimization, we restructure VPIC to avoid data transfers due to the high cost of data movement in modern systems. Data and simulation settings are set on the host CPU during initialization by default. Simulation data is then copied to the execution device and kept there throughout the simulation. Data movement between device and host is only done for communication and user diagnostics. On CPU-only platforms, where host and device are the same, data copies are eliminated automatically. This code structure ensures high performance on GPUs while Kokkos prevents any unnecessary data movement for CPU only systems.

Executing code in parallel introduces several tuning parameters for controlling how work is partitioned between the parallel threads of execution. The default settings determined by Kokkos achieve good performance and clean code. Still, performance can be further improved using hierarchical parallelism. Kokkos supports three levels of hierarchical parallelism: (1) thread teams, (2) individual threads, and (3) vector parallelism. In our second optimization, we apply hierarchical parallelism to the particle pusher as it is responsible for the majority of the computation. By explicitly defining the number of teams and threads per team, we exert more control over data access patterns and reuse. This optimization improves performance for the particle pusher and is easily applied to multiple target architectures.

Particle's sorting is a performance optimization in VPIC that significantly speeds up the particle pusher. The particle pusher must load interpolation data based on the particle's cell index. Current accumulation is also based on the cell index. Each cell must accumulate the generated current from all particles inside the cell. Thus keeping particles sorted based on cell index improves cache reuse. This feature is applied for CPU architectures as the GPU memory hierarchy requires a change in sorting order. The traditional sorting order results in many threads trying to access the same address which prevents the GPU from using the full memory bandwidth. In our third optimization, we address this issue by sorting the particles such that threads will access memory contiguously (e.g.,thread 0 accesses entry 0 and thread 1 accesses entry 1). This implementation not only improves the memory access pattern but also reduces the number of simultaneous writes to the same memory location.

GPU memory and atomic operation performance varies greatly between different manufacturers and hardware generations. Our fourth and final optimization uses the interoperability capabilities of Kokkos to implement target specific thread team reductions on GPUs. In particular, we introduce specialized CUDA code using warp intrinsics to accelerate current accumulation and reduce the number of atomic writes. Performance improvements are hardware dependent. The specialized code provides significant performance improvements for GPUs.

%% file: performance.tex
\section{Performance and Portability Study}
\label{sec5}

Our study of the newly optimized VPIC 2.0 code's performance and portability is executed on a diverse group of hardware platforms. For the performance, we consider both weak and strong scaling. For the portability, we present insights on the effectiveness of VPIC to run across multiple  architectures.  

\subsection{Hardware Platforms}
\label{sec:platforms}
We present results from five GPU-based platforms, and four CPU-based platforms. The architectures are sorted chronologically based on the year of release. The platform configurations are as follows. 

\subsubsection{GPU Platforms}

\noindent\textbf{NVIDIA Pascal P100 (2016)} experiments are run on the Kodiak computer at Los Alamos National Laboratory.  This computer consists of 133 nodes each with an Intel Xeon E5-2695~v4 CPU with 256 or 512~GB of DRAM and four NVidia Tesla P100 GPUs on a Mellanox InfiniBand EDR Fat-Tree interconnect.
\\

\noindent\textbf{NVIDIA Volta V100 (2017)} experiments are run on the Sierra supercomputer at Lawrence Livermore National Laboratory.  Sierra is a 125 petaflop, IBM Power Systems AC922 hybrid architecture system comprised of 4320 nodes, each with two 44-core IBM Power9 processors running at 2.3--3.8~GHz with 256 GB of RAM and four NVIDIA V100 GPUs each with 16~GB of HBM2 DRAM.
\\

\noindent\textbf{NVIDIA Turing RTX 5000 (2018)} experiments use the RTX partition of the Frontera cluster at the Texas Advanced Computing Center. This partition consists of nodes containing two Intel Xeon CPU E5-2620 v4 running at 2.10~GHz with 128~GB of RAM. Nodes are connected by Mellanox Infiniband FDR at 56~Gb/s configured in a fat-tree topology with a 4:1 oversubscription. Each node also contains four NVIDIA Quadro RTX 5000 GPUs with 16~GB of GDDR6 DRAM.
\\

\noindent\textbf{AMD Vega Radeon VII (2019)} experiments are executed on a single node consisting of two AMD EPYC 7302 16-core processors with 64~GB of DRAM and four AMD Radeon~VII GPUs with 16~GB of HBM2 DRAM each. \\

\noindent\textbf{NVIDIA Ampere A100 (2020)} experiments are completed on the NVIDIA DGX SuperPod \cite{SuperPodArch}, Selene, at NVIDIA. Each node within this system contains dual AMD EPYC 7742 CPUs with 64 cores per socket running at 2.25~GHz and 2~TB of DRAM. The nodes are connected together using Melanox Infiniband HDR at 200~Gb/s within a fat-tree topology that is configured for a 5:4 oversubscription. Each node contains eight NVIDIA A100 GPUs with 80~GB of HBM2 DRAM each.

\subsubsection{CPU Platforms}

\noindent\textbf{Intel Knights Landing (2016)} experiments use the KNL partition of the Trinitite testbed at Los Alamos National Laboratory. This partition consists of 100 nodes of a Cray XC40 system configured with a 3D Aries Dragonfly interconnect. Each node contains one 68~core Knights Landing processor with 16~GB of high-bandwidth MCDRAM and 96~GB of DRAM.
\\

\noindent\textbf{AMD Zen 2 (2019)} experiments are run on the Bell cluster at Purdue University \cite{PurdueCluster}. Each node contains two AMD EPYC 7662 64~core processors operating at 2.0~GHz and 256~GB of DRAM. Each node is connected with Mellanox Infiniband HDR at 100~Gb/s in a fat-tree configuration that is oversubscribed 3:1.   
\\

\noindent\textbf{Intel Cascade Lake (2019)} experiments are executed on the Frontera cluster at the Texas Advanced Computing Center~\cite{FronteraCluster}. Each node in Frontera contains two Intel Xeon Platinum 8280 processors with 28~cores operating at 2.70~GHz and 192~GB of memory. The nodes are connected with Mellanox Infiniband HDR at 100~Gb/s. The interconnect is configured in a fat-tree topology with an oversubscription factor of 11:9.
\\

\noindent\textbf{Fujitsu A64FX (2020)} experiments were ran on the Ookami cluster at Stony Brook University. Ookami is an HPE Apollo 80 system with 174 nodes. Each node contains a Fujitsu A64FX Arm processor~\cite{10.5555/3433701.3433763} with 32~GB of high-bandwidth memory and a 512~GB SSD. Nodes are connected in a 2-level tree using non-blocking HDR-200.

\subsection{Weak Scaling} 
\label{sec:weak_scaling}

For a test problem to study weak scaling, we adapt a large 3D simulation of stimulated Brillouin scattering (SBS) in a single laser speckle~\cite{albright2016multi} relevant for inertial confinement fusion experiments. Running the full simulation to completion would be prohibitive in our scaling tests, so we modify the simulation to be much shorter while preserving the numerical properties of the full simulation. Instead of a single very long laser pulse with a slow ramp in intensity, we use two counter-propagating laser pulses that ramp up linearly in field strength (quadratically in intensity) from zero at the start to twice the intensity of \cite{albright2016multi} at the end of the simulation. The simulation is stopped just before the two pulses collide. This arrangement gets particles moving near the start of the simulation, and thus is computationally similar to the majority of a full length simulation, and less like the uninteresting uniform cold plasma of early time steps.
    
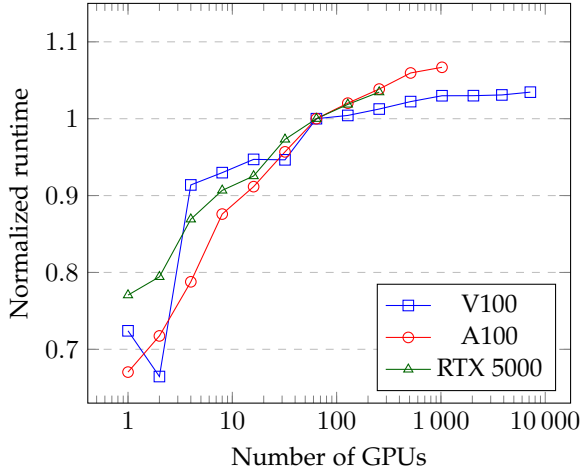
\begin{figure} [!h]

\centering   
\begin{tikzpicture}
\begin{semilogxaxis}[
    xlabel={Number of GPUs},
    ylabel={Normalized runtime},
    ymax=1.15,
    ymin=0.63,
    legend pos=south east,
    enlarge x limits=0.1,
    ymajorgrids=true,
    grid style=dashed,
    x tick label style={/pgf/number format/1000 sep=\,},
    log base 10 number format code/.code={%
        $\pgfmathparse{10^(#1)}\pgfmathprintnumber{\pgfmathresult}$%
},
]
        \addplot[
    y filter/.code={\pgfmathparse{\pgfmathresult/602.017}\pgfmathresult},
    color=blue,
    mark=square,
    ]
    coordinates {
        (1,435.899)(2,400.095)(4,550.147)(8,559.766)(16,570.147)(32,569.803)(64,602.017)(128,604.562)(256,609.584)(512,615.402)(1024,619.983)(2048,620.044)(3840,620.621)(7200,622.803)
    };
    
    %DGX runs
\addplot[
    y filter/.code={\pgfmathparse{\pgfmathresult/741.919}\pgfmathresult},
    color=red,
    mark=o,
    ]
    coordinates {
    (1,497.36)(2,532.273)(4,584.493)(8,649.86)(16,676.328)(32,709.911)(64,741.919)(128,756.914)(256,770.51)(512,786.051)(1024,791.531)
    };
    
%RTX
\addplot[
    y filter/.code={\pgfmathparse{\pgfmathresult/900.8855}\pgfmathresult},
    color=green!40!black,
    mark=triangle,
    ]
    coordinates {
    (1,694.1512)(2,715.5876)(4,783.057)(8,816.8672)(16,833.7366)(32,876.6388)(64,900.8855)(128,917.6254)(256,932.06)
    };

    \legend{V100, A100, RTX 5000}
\end{semilogxaxis}
\end{tikzpicture}

\caption{Weak Scaling on three GPU-based platforms (i.e., Sierra with Power 9 CPUs and V100 GPUs, Selene with EPYC 7742 CPUs and A100 GPUs, and the Frontera with Xeon E5-2620 v4 CPUs and RTX 5000 GPUs). Runtime is normalized to the 64 GPU runs, below which boundaries cause a reduction in per GPU communication.}
\label{fig:multi_arch_gpu}
\end{figure}

We simulate a fully ionized nitrogen plasma of temperatures $T_e=600$~eV and $T_i=150$~eV, with density $n_e=0.05n_{cr}$, where $n_{cr}=m_e \varepsilon_0\omega_L^2 / e^2$ is the critical density for a laser with angular frequency $\omega_0=2\pi c / \lambda_L$.
The plasma fills a simulation volume of \SI{80x20x20}{\micro\meter}.
Two identical laser pulses are incident on the plasma from the positive and negative $x$ direction.
Each pulse is a focused Gaussian beam of wavelength $\lambda_L=527$~nm and beam waist $w_0 = $\SI{4}{\micro\meter}, focused at the center of the simulation volume.
The pulses increase linearly in field strength over time to a maximum intensity $I_0=5 \times 10^{15}$~W/cm$^2$ at the end of the simulation, $t_\mathrm{stop}=(\mathrm{\SI{80}{\micro\meter}}/2)/c$.
The grid has resolution 6000x1500x1500, for a cell size very near the Debye length, and a domain decomposition of 60x15x15 across 13,500 GPUs.
Tests smaller than this full scale simulate a proportional fraction of the volume and use only one laser pulse.
Tests using 32 or fewer GPUs have an adjusted number of particles per cell to account for boundary conditions (there is a ``buffer" region at the simulation boundary where particles cannot be placed to separate particle and field boundary conditions), making the maximum number of particles a rank has at initialization equal for all tests. We run the weak  scaling  on  three cutting-edge, GPU-based platforms (i.e., Sierra with Power 9 CPUs and V100 GPUs, Selene with EPYC 7742 CPUs and A100 GPUs, and the Frontera with Xeon E5-2620 v4 CPUs and RTX 5000 GPUs). The reported runtime is measured on up 7,200 GPUs on Sierra, 1,024 on Selene, and 256 on Frontera; the times are on to normalized to 64 GPU runs. Fig.~\ref{fig:multi_arch_gpu} demonstrates the results from the three GPU-based platforms and their different GPUs, spanning both commercial and consumer grade cards. Runtime measurements for Frontera (RTX 5000) are the average of five runs with a standard deviation of 2.268. VPIC runtime in general is very consistent. Sierra (V100) and Selene (A100) measurements are from a single run. 
The weak scaling results show excellent performance across the GPU architectures considered. The best scaling performance is shown by the NVIDIA Volta (V100) hardware on the Sierra supercomputer. The tests on the Sierra's  7,200 GPUs  show only a 3.45\% slowdown versus the 64 GPU run. This is a testament to the capability of VPIC 2.0 to preserve performance while gaining in portability. 
Other the platforms (i.e., Selene and Frontera) scale about half as well as Sierra on the configurations that were made available to us for our testing.

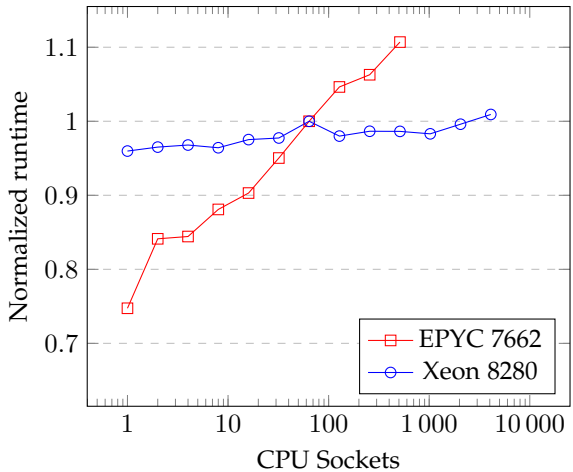
\begin{figure}[!h]

\centering   
\begin{tikzpicture}
\begin{semilogxaxis}[
    xlabel={CPU Sockets},
    ylabel={Normalized runtime},
     xmax=10000,
     ymin=0.66,
     ymax=1.11,
    legend pos=south east,
    enlarge x limits=0.1,
    enlarge y limits=0.1,
    ymajorgrids=true,
    grid style=dashed,
    x tick label style={/pgf/number format/1000 sep=\,},
    log base 10 number format code/.code={%
        $\pgfmathparse{10^(#1)}\pgfmathprintnumber{\pgfmathresult}$%
}
]
%Rome
\addplot[
    y filter/.code={\pgfmathparse{\pgfmathresult/2258.182}\pgfmathresult},
    color=red,
    mark=square,
    ]
    coordinates {
    (1,1687.722)(2,1899.608)(4,1906.62)(8,1989.216)(16,2039.226)(32,2146.016)(64,2258.182)(128,2362.832)(256,2400.1025)(512,2499.765)
    };
        
    \addplot[
    y filter/.code={\pgfmathparse{\pgfmathresult/2683.07625}\pgfmathresult},
        color=blue,
        mark=o,
        ]
        coordinates {
        (1,2575.068333)(2,2589.456667)(4,2596.923333)(8,2587.075)(16,2616.6025)(32,2622.4425)(64,2683.07625)(128,2629.312)(256,2646.85)(512,2646.374)(1024,2637.57)(2048,2672.406)(4096,2707.668)
        };

    \legend{EPYC 7662, Xeon 8280 }
\end{semilogxaxis}
\end{tikzpicture}

\caption{Weak scaling on two CPU-based platforms (i.e., Bell cluster at Purdue University using EPYC 7662 CPUs and the Frontera cluster with Xeon 8280 CPUs only).  Runtimes are normalized to the 64 CPU runs to match the GPU tests.}
\label{fig:multi_arch_cpu}
\end{figure}

We demonstrate the ability of VPIC 2.0 to run on CPU-based architectures for two platforms: the Bell cluster at Purdue University using EPYC 7662 CPUs and the Frontera cluster with Xeon 8280 CPUs only (without the use of GPUs). Fig.~\ref{fig:multi_arch_cpu}  shows the weak  scaling  on the two platforms. The runtime values are the average of five runs and are normalized to the 64 CPU runs, as in the GPU-based testing. The presented runtimes are the average of five runs. For our tests, we run multiple  MPI  ranks  per  socket, thus  the  per  socket  communication  is  constant  for  4  or more sockets.
While these performance results are notably slower than the GPU runs in absolute terms, they still add a lot of value from the perspective of our portability study. Furthermore, as accelerators become the driving architectures for large scale simulations, the community gathering around VPIC is expected to use GPU-based platforms rather than CPU-only ones.

\subsection{Strong Scaling}
\label{sec:strong_scaling}

The cache increase on the A100 to 40~MB from 6~MB on the V100 has tangible implications for how these two cards are best utilized in VPIC 2.0. A significant part of the particle push is depositing the current generated from particle movement to grid points. VPIC 2.0 periodically sorts the particles by voxel such that the number of grid points being written to is minimal at any given time. Minimizing the number of grid points to write to, allows the grid points to be stored in cache, resulting in faster writes. The A100 raises the intriguing possibility of storing the entire grid in cache, obviating the particle sort, and possibly producing super-linear speedup behavior in a strong scaling test as the total available cache increases.

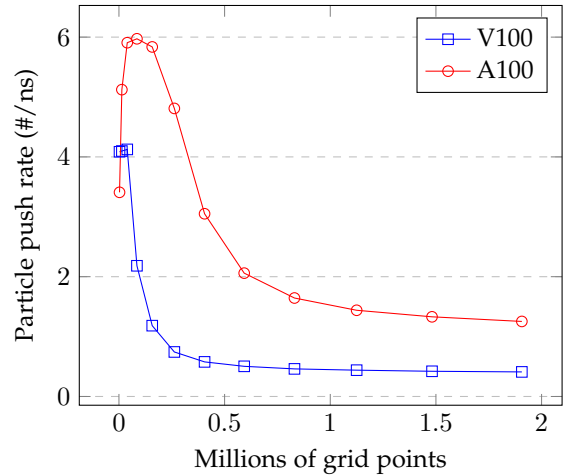
\begin{figure}[!h]
    \centering
\begin{tikzpicture}
\begin{axis}[
    xlabel={Millions of grid points},
    ylabel={Particle push rate (\#/ns)},
    legend pos=north east,
    ymajorgrids=true,
    grid style=dashed,
]

\addplot[
    color=blue,
    mark=square,
    ]
    coordinates {
        (0.002744,4.085645340449165)
(0.013824,4.098150709492342)
(0.039304,4.125146636071829)
(0.085184,2.1820661438799864)
(0.157464,1.182382500738989)
(0.262144,0.7432181345224824)
(0.405224,0.5773776230084983)
(0.592704,0.5035167497993801)
(0.830584,0.46056419113413927)
(1.124864,0.4392708104546453)
(1.481544,0.4207426107078995)
(1.906624,0.41013547287338353)

    };
    \addplot[
    color=red,
    mark=o,
    ]
    coordinates {
        (0.002744,3.4086067319982956)
(0.013824,5.12273212379936)
(0.039304,5.9064807219032)
(0.085184,5.972130059721301)
(0.157464,5.836102780254519)
(0.262144,4.809619238476953)
(0.405224,3.0507181899072076)
(0.592704,2.0593198524154106)
(0.830584,1.6439857521234815)
(1.124864,1.4392228196773742)
(1.481544,1.3301311657121744)
(1.906624,1.2528493622648733)

    };
    \legend{V100,A100}

\end{axis}
\end{tikzpicture}
    \caption{Number of particle pushes per nanosecond as a function of grid size on Sierra with V100 and Selene with A100. The total number of particles is held constant and all simulation time not in the particle push is excluded.}
    \label{fig:pushVgrid}
\end{figure}

We investigate this feature with tests similar to the weak scaling tests above, but with particle sorting turned off and only using the two larger supercomputers (i.e., Sierra with the V100 and Selene with the A100). The number of particles is kept constant while grid size varies. Fig.~\ref{fig:pushVgrid} shows the number of  particle  pushes  per  nanosecond  as  a function of grid size. The figure confirms that there is indeed a grid size where particle push performance increases dramatically, both for V100 and A100. Performance peaks for the V100 with approximately 4 particle pushes per nanosecond at 13,824 grid points. On the other hand, the A100 reaches 6 particle pushes per nanosecond with a grid size of 85,184. Further more, the performance jump occurs at roughly 6 times more grid points for A100 vs. V100, similar to the increase in cache. If this increase is from cache effects, we should see super-linear speedup in the following strong scaling tests. We suspect the performance drop for very few grid points, and thus extremely high particles per cell, on A100 is a result of colliding atomic writes in the current deposition.

Carefully selecting the size of our grid to match the peak performance in Figure~\ref{fig:pushVgrid}, we run strong scaling tests on Sierra with V100 GPus (i.e., from 1 to 32 GPUs), as shown in Fig.~\ref{fig:NVIDIA_Volta_strong}, and on Selene with A100 GPUs (i.e., from 8 to 512 GPUs), as shown in Fig.~\ref{fig:NVIDIA_Ampere_Strong}.
As suggested by Fig.~\ref{fig:pushVgrid}, we do indeed observe super-linear scaling for both architectures.
For V100, we observe a 25x speedup for an 8x increase of GPUs, from 1 to 8. Beyond 8, the GPUs are very empty and communication overhead starts to cancel out the super-linear speedup. On A100, we see a 19x speedup for an 8x increase of GPUs, from 8 to 64. Unlike the V100, we see near-ideal scaling all the way to 512 GPUs, the largest allocation we were able to test.
\begin{figure} [!h]
    \centering
    \begin{tikzpicture}
\begin{axis}[
    xlabel={Number of GPUs},
    ylabel={Runtime (s)},
    ymin=8,
    xmax=100,
    ymode=log,
    xmode=log,
    legend pos=north east,
    ymajorgrids=true,
    enlarge x limits=0.1,
    grid style=dashed,
        x tick label style={/pgf/number format/1000 sep=\,},
    log base 10 number format code/.code={%
        $\pgfmathparse{10^(#1)}\pgfmathprintnumber{\pgfmathresult}$%
    }
]

\addplot[
    color=blue,
    mark=square,
    ]
    coordinates {
    (1,1254.61)(2,443.284)(4,140.137)(8,49.9993)(16,34.1906)(32,30.137)
    };
\addplot[
    color=black,
    mark=x,
    ]
    coordinates {
    (1,1254.61)(2,1254.61/2)(4,1254.61/4)(8,1254.61/8)(16,1254.61/16)(32,1254.61/32)
    };    
    \legend{V100, Ideal}
    
\end{axis}
\end{tikzpicture}

    \caption{Strong scaling on the Sierra with V100 GPUs. As the number of GPUs increases, more of the grid is kept in cache; as a result we get super-linear speedup. Communication costs eventually catch up with more than 16 GPUs.}
    \label{fig:NVIDIA_Volta_strong}
\end{figure}
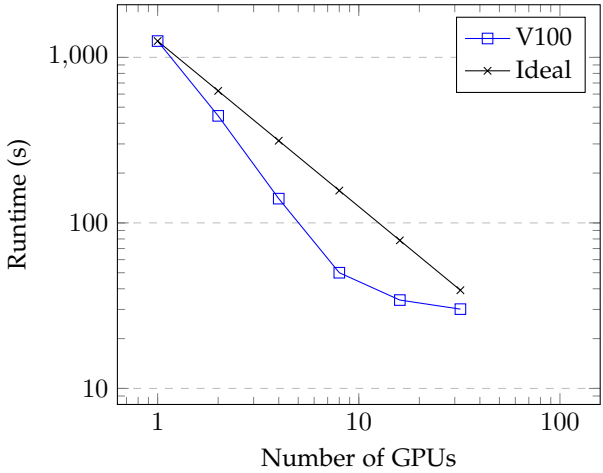
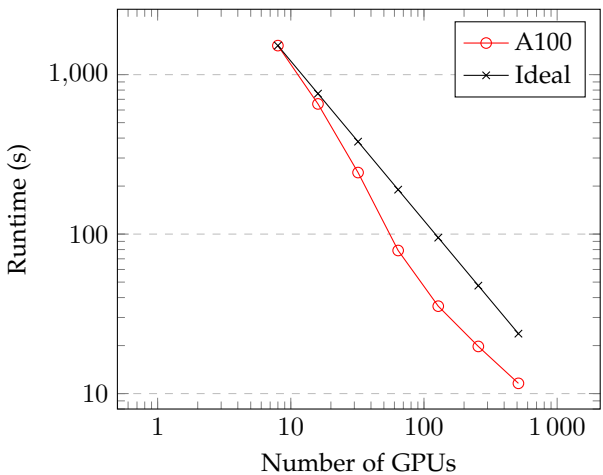
\begin{figure}

\centering   
\begin{tikzpicture}
\begin{semilogxaxis}[
    xlabel={Number of GPUs},
    ylabel={Runtime (s)},
    xmin=1, xmax=1050,
    ymode=log,
    ymin=8,
    enlarge x limits=0.1,
    legend pos=north east,
    ymajorgrids=true,
    grid style=dashed,
    x tick label style={/pgf/number format/1000 sep=\,},
    log base 10 number format code/.code={%
        $\pgfmathparse{10^(#1)}\pgfmathprintnumber{\pgfmathresult}$%
    }
]
\addplot[
    color=red,
    mark=o,
    ]
    coordinates {
    (8,1520.41)(16,655.257)(32,243.194)(64,78.928)(128,35.3268)(256,19.7548)(512,11.5922)
    };
\addplot[
    color=black,
    mark=x,
    ]
    coordinates {
    (8,1520.41)(16,1520.41/2)(32,1520.41/4)(64,1520.41/8)(128,1520.41/16)(256,1520.41/32)(512,1520.41/64)
    };
 \legend{A100,Ideal}
\end{semilogxaxis}
\end{tikzpicture}

\caption{Strong scaling on the Selene with A100 GPUs. The A100 GPUs maintain super-linear speedup with more GPUs than the V100 test thanks to the larger cache.}
\label{fig:NVIDIA_Ampere_Strong}
\end{figure}

The super-linear speedup from storing the grid in the cache has practical applications for certain classes of problems. Simulations with fewer grid points or high particle per cell requirements (e.g., cross-beam energy transfer inertial confinement fusion simulations~\cite{yin2019saturation}) can cut their runtime down significantly.

Smaller, non-full-scale simulations also benefit significantly from the super-linear speedup. The faster runtime enables domain experts and developers to quickly iterate through test parameters and tune their simulations before moving on to full-scale runs.

\subsection{Portability Study}

Our portability study assesses how effectively VPIC 2.0, a single source code, can run across multiple architectures. To show to achieved portability, we present performance of each architecture listed in Sec.~\ref{sec:platforms} for VPIC 2.0.
 
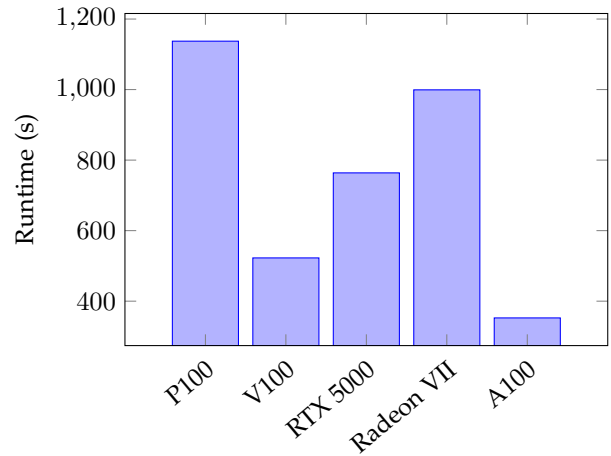
\begin{figure} [!ht]
    \centering
    \begin{tikzpicture}
    \begin{axis}[
        ylabel={Runtime (s)},
        ybar stacked,
        height=6cm,
        bar width=25pt,
        enlarge x limits=0.25,
        legend pos=north west,
        xtick={data},
        xticklabels={P100, V100, RTX 5000, Radeon VII, A100},
        x tick label style={align=center, rotate=40, anchor=north east},
        scaled y ticks=false,
        axis on top,
        ]
    \addplot+[every node near coord/.style=black] plot
        coordinates {
         (1, 1137.25) (2, 522.654) (3,763.7598) (4, 999.217) (5, 352.5004) 
        };
    
    \end{axis}
    \end{tikzpicture}

    \caption{Total simulation time for small scale GPU performance (8 cards). Results are ordered chronologically (based on the GPU releases).}
    \label{fig:single_gpu}
\end{figure}

Fig.~\ref{fig:single_gpu} shows a comparison of the different runtime measurements in seconds for the GPU-based platforms when running on 8 GPUs, with 1 MPI rank per GPU. The total simulation time is shown and is predominately compute time, as communication time is minimal at 8 MPI ranks.
For the NVIDIA based platforms (i.e., P100, V100, RXT 5000, and A100) we see a large jump between P100 and V100, which can be explained by the increased atomic operation support in the V100. We attribute the further jump from V100 to A100 to the increase in both cache size and FP32 performance. 
The RTX 5000, while slower than the V100, performs admirably for a workstation card and proves itself very capable of single-precision scientific simulations. The AMD Radeon VII underperforms when compared to the RTX 5000, despite having similar theoretical compute performance and superior memory bandwidth. The difference in performance can be attributed to two factors: (1) the maturity of the Kokkos AMD backend; and (2) the fact that to date NVIDIA cards have been used as the development and testing platform for this work. The performance of VPIC 2.0 may further increase as Kokkos improves the AMD back-end.

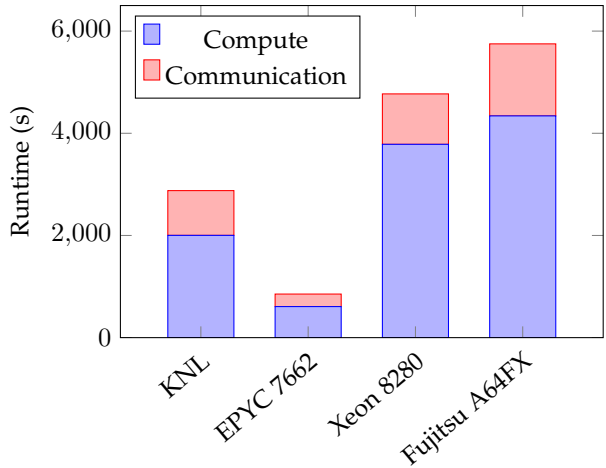
\begin{figure}[!ht]
    \centering
    \begin{tikzpicture}
    \begin{axis}[
        ylabel={Runtime (s)},
        ybar stacked,
        bar width=25pt,
        enlarge x limits=0.25,
        ymin=0,
        ymax=6500,
        height=6cm,
        legend pos=north west,
        legend style={
        },
        xtick={data},
        xticklabels={KNL, EPYC 7662, Xeon 8280, Fujitsu A64FX},
        x tick label style={align=center, rotate=40, anchor=north east},
        scaled y ticks=false,
        axis on top,
        ]
    \addplot+[every node near coord/.style=black] plot
        coordinates {
        (1, 2001.83) (2,607.8468) (3, 3783.9) (4, 4.34E+03)
        };
    
    \addplot+[every node near coord/.style=black] plot
        coordinates {
        (1, 8.76E+02) (2,245.16) (3, 986.85) (4, 1.41E+03)
        };
    
    \legend{\strut Compute, \strut Communication}
    \end{axis}
    \end{tikzpicture}

    \caption{Compute and communication time for small scale CPU performance (8 sockets). Results are ordered chronologically (based on the CPU releases).}
    \label{fig:single_cpu}
\end{figure}

Fig.~\ref{fig:single_cpu} provides a comparison of CPU-based platforms, including chips from AMD, Intel, and ARM/Fujitsu. For test normalization we compare 8 sockets of each platform, which represents a regime where MPI costs are non-negligible. While it is an impressive feat that the same source code of VPIC 2.0 can run on such a diverse range of hardware, it is interesting to note that the AMD Zen 2 is the only platform that achieves performance comparable to that shown in the GPU study. We attribute the lower performance on CPU-based platforms to the limited vectorization support in the current version of Kokkos (version 3.3). For further optimizations, one could formulate the VPIC 2.0 code expression that is better able to vectorize. We do not address vectorization in this work as that introduces non-trivial code divergence, and the focus of this study is on achieving portability over platform specific performance. It is our belief that for cutting-edge performance on a specific target platform, some level of specialization will always be required. In many cases the portability abstraction layer may be able to achieve this, but other times the burden will fall to the programmer. 

With a portability study such as ours that relies on an underlying toolkit to facilitate the codes operation on a variety of platforms, it is somewhat inevitable that the performance results reflect the priorities and maturity of the framework, as well as reflecting the facets of the target hardware. It is clear from these results that the code, as written in its present state, maps best to GPU platforms. Proposals have been made to extend Kokkos SIMD support, and we look forward to being able to leverage these improvements for VPIC 2.0.

%% file: usecases.tex
\section{Scientific Impacts of VPIC 2.0}
\label{sec6}

The ability for any code to readily deploy itself on a new machine, with limited manual refactoring effort, is a huge win that actively drives down the time to scientific discovery. As the heterogeneity of modern platforms continues to increase to meet the demands of exascale, effective use of these platforms translates directly to codes such as VPIC being able to run simulations at unprecedented scales, and opens the door to performing analysis and investigation which was previously intractable. Sec.~\ref{sec:weak_scaling} demonstrates that VPIC 2.0, by following the methodology laid out in this paper, is well prepared for the exascale challenge. 

While enabling portability, VPIC 2.0 is also able to assure performance. The performance-portability trade-off allows simulations to complete quickly and without multiple restarts by using accelerators and emerging architectures; such platforms are not optimized or supported by the original VPIC code (VPIC 1.0). The reduction in execution costs when using accelerators, coupled with the portability presented in this paper, facilitate the following changes in VPIC simulations:
\begin{itemize}
    \item Simulating more timesteps in fewer core-hours enables simulations which advance further in time, allowing us to set our sights towards performing longer fully resolved electromagnetic simulations than were previously possible.
    \item Advanced and expensive diagnostics can now be performed within the timestep that allows for more accurate analysis. The extensive analysis enables new fundamental insights to particle acceleration modeling at unprecedented scales~\cite{guo2019determining}.
    \item Previous simulations, which took considerable resources, can now be run multiple times to analyze stochastic variation in, for example, short-pulse laser experiments, which in turn can generate sufficient data for machine learning analysis.
\end{itemize}
The newly release VPIC code opens up new avenues of science, carrying the domain of plasma physics research to exascale level simulations

%% file: related_work.tex
\section{Related Work}
\label{sec7}

While the computational characteristics of PIC codes are fairly well understood, much of the communities optimization and development efforts to date have been focused around platform and hardware specific optimization~\cite{surmin2016co, nakashima2015manycore}.
As heterogeneity within leading supercomputers increases~\cite{khan2021analysis}, it is no longer viable to periodically optimize for a single target platform, making portability more important than ever. Within the PIC community, portability is under-explored, and our work seeks to address this weakness. Example projects such as PIConGPU~\cite{burau2010picongpu,zenker2016performance} and the AMITIS project~\cite{fatemi2017amitis} demonstrate the applicability of the PIC algorithm to GPU architectures, but do not make a concerted effort to address the portability issues presented by the increasing diversity of modern HPC platforms.

We leverage the Kokkos framework to address many aspects of the performance-portability trade-off. While Kokkos represents a fairly modern approach to code portability, it has already established itself as a key technology in the race for performance portability, and has successfully helped a variety of applications run on multiple large-scale platforms. These target applications span science domains including finite element analysis~\cite{demeshko2019toward}, computational fluid dynamics~\cite{eichstadt2020comparison}, and even deep neural networks~\cite{ellis2019scalable}. It is of particular note previous work by the molecular dynamics community to leverage Kokkos for performance portability \cite{gayatri2020rapid, pennycook2018evaluating, womeldorff2017taking}. Molecular dynamics is an important workload for Sandia National Laboratory, the creators of Kokkos; and is also the best explored workload that shares computational similarities with the PIC method. Both methods are particle dominated, and rely on the fast processing of large amounts of particles to simulate the kinetic level behavior of a large system.
It is important to note, however, that Kokkos is not the only project which seeks to address issues of performance-portable code development. These tools can be split into two types: high-level frameworks (such as Kokkos, RAJA~ \cite{hornung2014raja}, or HPX~\cite{kaiser2020hpx})  and lower-level tools for which code can be generated (such as OpenMP, SYCL, or HIP).
A comparison of these technologies has already been the subject of previous studies~\cite{martineau2015performance, kirk2017achieving}. Our selection of Kokkos to support VPIC's portability is based on the fact that Kokkos allows for high-level programmability while still offering good low-level performance.

%% file: conclusion.tex
\section{Conclusion}
\label{sec8}

In this work we present a performance-portable version of VPIC, a kinetic Particle-in-Cell code that has been used to do some of the worlds largest plasma physics simulations. This version of VPIC (VPIC 2.0) relies on the Kokkos framework for portability. We capture the VPIC 2.0 performance on nine different platforms (including three of the top 10 Top500 systems)  and scale our tests  on up to $7,200$ GPUs. We demonstrate super-linear strong scaling on the NVIDIA A100 architecture and provide insights on the scientific impacts of VPIC 2.0 on scientific discovery for the next generation exascale supercomputers.